Tensor Formulation of the General Linear Model with Einstein Notation

Abbreviated Title: **Tensor Formulation of the GLM**

Gavin T. Kress, MS[*]

*Corresponding Author*

**Author of Correspondence:** Gavin Kress, MS, **Email:** gavinkress@gmail.com, **Tel:** (865) 804-2847, **Postal Address:** 309 E 108[th] Street, Unit 3A New York, New York 10029

**Equation Count:** 28, **Word Count:** 179 (Abstract), 1582 (Main Text)

**Author Contributions:**

**Gavin T. Kress** (ORC ID: 0000-0001-5152-1170, gavinkress@gmail.com) – **Conceptualization, Formal Analysis, Investigation, Methodology, Visualization, Writing – Original Draft Preparation, Writing – Review & Editing**

**Funding Source:** This research received no specific grant from any funding agency in the public, commercial, or not-for-profit sectors

**Conflict of Interest Disclosure:** None

**Acknowledgments:** None

# Tensor Formulation of the General Linear Model with Einstein Notation

The general linear model is a universally accepted method to conduct and test multiple linear regression models. Using this model, one has the ability to simultaneously regress covariates among different groups of data. Moreover, there are hundreds of applications and statistical tests associated with the general linear model.

However, the conventional matrix formulation is relatively inelegant which yields multiple difficulties including slow computation speed due to a large number of computations, increased memory usage due to needlessly large data structures, and organizational inconsistency. This is due to the fundamental incongruence between the degrees of freedom of the information the data structures in the conventional formulation of the general linear model are intended to represent and the rank of the data structures themselves.

Presented here is an elegant reformulation of the general linear model which involves the use of tensors and multidimensional arrays as opposed to exclusively flat structures in the conventional formulation. To demonstrate the efficacy of this approach, a few common applications of the general linear model are translated from the conventional formulation to the tensor formulation.

## Introduction

The general linear model (GLM) or general multivariate regression model is a widely accepted technique across multiple fields to perform several multiple linear regression models. It offers advantages such as the ability to simultaneously regress covariates among different groups of data, among others. The applications and statistical tests derived from and expressed using the conventional matrix formulation of the GLM are numerous and multifaceted (Khuri et al., 2006; Mccullagh, 1984).

However, the conventional matrix formulation is relatively inelegant in some embodiments, yielding compromised computational efficiency and increased order of complexity in automation of statistical tests. For example, in cases in which multiple groups are modeled, the matrix formulation lacks the dimensionality to encode the relevant linear coefficients and variables. The brute force solution to this in the conventional formulation is to simply stagger the indices corresponding to the various groups such that the relevant parameters and variables are all encoded in a sparse, flat data structure.

Put forth here is an elegant reformulation of the GLM, such that the data structures describing the important parameters and variables are tensors represented in Einstein notation (Åhlander, 2002). To demonstrate the efficacy of this approach, a general description of the formulation will precede a few brief examples of applications for which this formulation is more elegant than the matrix formulation.

## Conventional Formulation of the GLM

The GLM most generally consists of N domain variables from which a linear atlas is generated which maps this domain space of $\mathbb{R}^N$ to a linear manifold in $\mathbb{R}^1$ defined by an outcome variable. Moreover, the parameters defining such atlas depend on the group from which the domain variables are derived. This linear atlas is conventionally expressed as described in **Eq. 1**.

$$y = X\beta \tag{1}$$

Where y is the outcome variable, $X$ is the covariant vector consisting of the domain variables with the first entry equal to one corresponding to the intercept, and $\beta$ is the contravariant vector consisting of the coefficients of said domain variables. This map is generated with a function of a series of residuals defined in **Eq. 2**.

$$Y = X\beta + N \tag{2}$$

In this, $Y$ is the contravariant vector representing samples of the linear outcome manifold in $\mathbb{R}^1$ for a particular group, while $X$ is a matrix describing a series of the same covariant vectors in **Eq. 1**, which were experimentally determined to map to said samples of the outcome manifold. By choosing the parameters of $\beta$, the linear map will, at best, approximate the experimentally defined atlas, implying the existence of residuals $N$.

The matrices are written out explicitly for only one regressor or domain variable in **Eq. 3**.

$$\begin{bmatrix} Y^0 \\ Y^1 \\ Y^2 \\ \vdots \end{bmatrix} = \begin{bmatrix} 1 & X^0 \\ 1 & X^1 \\ 1 & X^2 \\ \vdots & \vdots \end{bmatrix} \begin{bmatrix} b \\ m \end{bmatrix} + \begin{bmatrix} n^0 \\ n^1 \\ n^2 \\ \vdots \end{bmatrix} \tag{3}$$

At this point, the model is inconspicuously inelegant. However, upon the introduction of numerous groups from which the experimental data is sampled, it becomes evident. Expanding this example to one regressor in two groups is shown in **Eq. 4**.

$$\begin{bmatrix} Y^{1,0} \\ Y^{1,2} \\ \vdots \\ Y^{2,0} \\ Y^{2,1} \\ \vdots \end{bmatrix} = \begin{bmatrix} 1 & 0 & X^{1,0} & 0 \\ 1 & 0 & X^{1,1} & 0 \\ \vdots & 0 & \vdots & 0 \\ 0 & 1 & 0 & X^{2,0} \\ 0 & 1 & 0 & X^{2,1} \\ 0 & \vdots & 0 & \vdots \end{bmatrix} \begin{bmatrix} b^1 \\ b^2 \\ m^1 \\ m^2 \end{bmatrix} + N \tag{4}$$

Clearly in this staggered configuration, as more groups are introduced into the model, the matrix continues to grow, with the majority of entries being equal to zero, which yields an unnecessarily large number of computations and quantity of memory usage. Moreover, without a priori knowledge of both the number of groups and number of regressors, it is impossible to predict the organizational structure of the matrix. In general, a model with $m$ regressors, $n$ groups, and $k$ data points in each group will require a matrix $\mathbf{X}$ as a series of $k \times n$ covariant vectors, each with $n$ indices encoding only 1s and 0s which are implicated as coefficients of the intercepts and $m \times n$ indices encoding the experimental independent data. The resultant size of the data structure in question is then $kn^2(m+1)$.

### Tensor Formulation of the GLM

These problems are circumvented with an alternative tensor formulation of the model expressed in Einstein notation. Notably, **Eq. 2** Can be written as **Eq. 5**.

$$Y^{k,\lambda} = X_\alpha^{k,\lambda} \beta^{\alpha,\lambda} + N^{k,\lambda} \tag{5}$$

Here, k indexes the samples of the experimental mapping, $\lambda$ indexes over the group, and $\alpha$ indexes over the intercept and each regression variable or parameter. From this, the extension to the translation of **Eq. 1** is trivial. An example of such formulation in hybrid matrix Einstein notation with one regression parameter is shown in **Eq. 6**.

$$\begin{bmatrix} Y^{0,\lambda} \\ Y^{1,\lambda} \\ Y^{2,\lambda} \\ \vdots \end{bmatrix} = \begin{bmatrix} X_0^{0,\lambda} = 1 & X_1^{0,\lambda} \\ X_0^{1,\lambda} = 1 & X_1^{1,\lambda} \\ 1 & X_1^{2,\lambda} \\ \vdots & \vdots \end{bmatrix} \begin{bmatrix} b^\lambda \\ m^\lambda \end{bmatrix} + \begin{bmatrix} n^{0,\lambda} \\ n^{1,\lambda} \\ n^{2,\lambda} \\ \vdots \end{bmatrix} \tag{6}$$

### GLM Contrast Matrix in Tensor Notation

The null hypothesis ($\mathbf{H_0}$) statements to test in the GLM take the form of a linear combination of the atlas parameters in $\boldsymbol{\beta}$ is equal to 0. This linear combination is conventionally expressed in a manner outlined in **Eq. 7**.

$$\boldsymbol{g} = \boldsymbol{C\beta} \tag{7}$$

Where, $\mathbf{g}$ corresponds to the value $\mathbf{H_0}$ asserts is equal to zero and $\boldsymbol{C}$ is the GLM contrast matrix, which is a covariant vector with indices corresponding to the atlas parameters in $\boldsymbol{\beta}$ which serves as their linear coefficients in the $\mathbf{H_0}$ statement. This is expressed explicitly for 2 hypotheses ($g^0$ and $g^1$), one regressor, and two groups in **Eq. 8**.

$$\begin{bmatrix} g^0 \\ g^1 \end{bmatrix} = \begin{bmatrix} C_{b^1}^0 & C_{b^2}^0 & C_{m^1}^0 & C_{m^2}^0 \\ C_{b^1}^1 & C_{b^2}^1 & C_{m^1}^1 & C_{m^2}^1 \end{bmatrix} \begin{bmatrix} b^1 \\ b^2 \\ m^1 \\ m^2 \end{bmatrix} \tag{8}$$

Of note, even though, as it is shown here, the conventional formulation is compatible with multiple hypothesis testing, it has not been conventionally implemented in this way and each row of the vector **g** and matrix **C** are implemented as separate statements. The compatible expression in the tensor formulation is shown in **Eq. 9**.

$$g^\eta = C^\eta_{\alpha,\lambda} \beta^{\alpha,\lambda} \tag{9}$$

This model is more naturally compatible with multiple $\boldsymbol{H_0}$, which $\eta$ indexes over, for F-testing or multiple t-tests.

An example of such expression with two $\boldsymbol{H_0}$ statements in a model with two separate groups and a single regressor is outlined component-wise in **Eq. 10-23**.

$$\beta^{0,0} = b_1, \; \beta^{1,0} = m_1 \tag{10-11}$$

$$\beta^{0,1} = b_2, \; \beta^{1,1} = m_2 \tag{12-13}$$

$$C^0_{0,0} = C^0_{b_1}, \; C^0_{1,0} = C^0_{m_1} \tag{14-15}$$

$$C^1_{0,0} = C^1_{b_1}, \; C^1_{1,0} = C^1_{m_1} \tag{16-17}$$

$$C^0_{0,1} = C^0_{b_2}, \; C^0_{1,1} = C^0_{m_2} \tag{18-19}$$

$$C^1_{0,0} = C^1_{b_1}, \; C^1_{1,0} = C^1_{m_1} \tag{20-21}$$

$$g^0 = C^0_{b_1} b_1 + C^0_{m_1} m_1 + C^0_{b_2} b_2 + C^0_{m_2} m_2 \tag{22}$$

$$g^1 = C^1_{b_1} b_1 + C^1_{m_1} m_1 + C^1_{b_2} b_2 + C^1_{m_2} m_2 \tag{23}$$

While the number of elements in this expression remains unchanged in the reformulated version, the rank of the data structures are congruent with their degrees of freedom. Moreover, translating this expression is necessary to implement applications which use it and would benefit both from a computational speed and memory requirement perspective from the reformulation, such as the multiple t-tests application outlined below.

### GLM Multiple T-Test in Tensor Notation

The justification for representing the various applications of the GLM in a tensor formulation is self-evident at this point, and in most cases it is straightforward to infer such representations from the conventional notation. However, this is not always true, especially in embodiments which require inverting matrices.

The t-statistic is of the most important of these embodiments, which is represented in the conventional matrix notation in **Eq. 24**.

$$t = \frac{C\beta}{\sqrt{\sigma^2 C(X^T X)^{-1} C^T}} \quad (24)$$

Where the t-statistic is generated separately for each $H_0$, and $\sigma^2$ is the variance of the experimental outcome measure in most cases.

To express this in the tensor formulation, it is evident that the numerator is g, which is indexed for each hypothesis and, consequently, so is t. Moreover, it is clear that contracting a matrix with its transposed self can be expressed as shown in **Eq. 25**.

$$W_\alpha^{\alpha',\lambda} = \boldsymbol{X}^T\boldsymbol{X} = X_k^{\alpha',\lambda}X_\alpha^{k,\lambda} \qquad (25)$$

Moreover, the inverse of a matrix expressed in tensor notation is computed as Dr. Roger Penrose puts forth (Roger Penrose, 1971) and as is shown in **Eq. 26**.

$$[W^{-1}]_\mu^{\zeta,\lambda} = 2[\varepsilon^{\alpha_1,\alpha_2}\varepsilon_{\alpha'_1,\alpha'_2}W_{\alpha_1}^{\alpha'_1,\lambda}W_{\alpha_2}^{\alpha'_2,\lambda}]^{-1}\varepsilon^{\zeta,\alpha_2}\varepsilon_{\mu,\alpha'_2}W_{\alpha_2}^{\alpha'_2,\lambda} \qquad (26)$$

Where $\varepsilon$ is the totally antisymmetric Levi-Civita symbol which is defined from the sign by the permutation of its indices such that each value is a power of (-1) which matches the parity of the permutation, otherwise the value is zero. It is important to note that this is a general algorithm to invert tensors which scales in number of computations with the number of elements in said tensor, regardless of rank. Therefore, a tensor of rank 2 (e.g. **X** in the conventional formulation) would require more computations to invert than a rank 3 tensor that has fewer data elements (e.g. $X_\alpha^{k,\lambda}$ in the novel formulation). Extending this algorithm to $(\boldsymbol{X}^T\boldsymbol{X})^{-1}$ yields **Eq. 27**.

$$(\boldsymbol{X}^T\boldsymbol{X})^{-1} = 2[\varepsilon^{\alpha_1,\alpha_2}\varepsilon_{\alpha'_1,\alpha'_2}(X_k^{\alpha'_1,\lambda}X_{\alpha_1}^{k,\lambda})(X_k^{\alpha'_2,\lambda}X_{\alpha_2}^{k,\lambda})]^{-1}\varepsilon^{\zeta,\alpha_2}\varepsilon_{\mu,\alpha'_2}(X_k^{\alpha'_2,\lambda}X_{\alpha_2}^{k,\lambda}) \qquad (27)$$

∴

$$t^\eta = \frac{C_{\alpha,\lambda}^\eta \beta^{\alpha,\lambda}}{\sqrt{(\sigma^\eta)^2 C_{\alpha,\lambda}^\eta 2[\varepsilon^{\alpha_1,\alpha_2}\varepsilon_{\alpha'_1,\alpha'_2}(X_k^{\alpha'_1,\lambda}X_{\alpha_1}^{k,\lambda})(X_k^{\alpha'_2,\lambda}X_{\alpha_2}^{k,\lambda})]^{-1}\varepsilon^{\alpha,\alpha_2}\varepsilon_{\alpha',\alpha'_2}(X_k^{\alpha'_2,\lambda}X_{\alpha_2}^{k,\lambda})C_\lambda^{\alpha',\eta}}} \qquad (28)$$

**Results and Discussion**

The tensor formulation of the GLM drastically decreases the number of elements in the data structures and reduces the quantity of operations required to perform computations with said data structures, especially as more groups, regressors, and hypotheses are incorporated in the model. Specifically, a model which would require $kn^2(m+1)$ elements in the matrix **X** in the conventional formulation now only requires $knm$ elements in the corresponding reformulated data structure, thereby reducing the number of elements by a factor of $\frac{n(m+1)}{m}$. Depending on the data type used in the implementation, this formulation can significantly improve the memory required to store large data structures. Moreover, as the number of operations to perform a function scale with the size of a data structure this has the potential to significantly reduce the time required to test various hypotheses. Since the vast majority of the applications of the GLM require computations with the matrix **X**, these improvements are ubiquitous among them.

Additionally, the automation of hypothesis testing with the GLM is significantly simplified in the tensor formulation by the property that no a priori knowledge of the number groups, regressors, and hypotheses is needed to infer the structural organization of the data.

Finally, this solution is simply more elegant, as the rank of the tensors is complementary to the degrees of freedom of the information which the data structure in the GLM is designed to interact with.

There are hundreds of unique applications of the GLM, each of which can be formulated in this proposed manner. Presented here are the general structures of such formulations with a few examples, but the literature would benefit from further translation of other applications.